\documentclass{ws-procs9x6}            % default, citations in superscript
\begin{document}
\title{Phase transition via entanglement entropy in $AdS_3/CFT_2$  superconductors }

\author{Davood Momeni$^*$, Kairat Myrzakulov$^\dagger$, Ratbay Myrzakulov $^\star\star$}

\address{Eurasian International Center for Theoretical Physics and\\ 
Department of General \& Theoretical Physics,Eurasian National
University, Astana 010008, Kazakhstan.\\
$^*$E-mail: momeni\_d@enu.kz\\
$^\dagger$E-mail:myrzakulov\_kr@enu.kz 
\\
$^\star\star$E-mail: rmyrzakulov@gmail.com
\\
www.enu.kz}

%\author{A. N. Author}

%\address{Group, Laboratory, Street,\\
%City, State ZIP/Zone, Country\\
%E-mail: an\_author@laboratory.com}

\begin{abstract}
The purpose of this report is to provide a framework for defining  phase transition processes in two dimensional holographic superconductors, and to illustrate how they are useful to be described by holographic entanglement entropy. 
We study holographic entanglement entropy in a two dimensional fully backrected model for holographic superconductors. We prove that phase transition could be observe using a discontinuty in the first order of entropy.
\end{abstract}

\keywords{AdS/CFT Correspondence; Holography and
condensed matter physics (AdS/CMT); Entanglement entropy.}

\bodymatter

\section{Introduction}   
Entanglement lies at the heart of many aspects of quantum information theory and it is therefore desirable to understand its structure as well as possible. One attempt to improve our understanding of entanglement is the study of our ability to perform information theoretic tasks locally based on Anti-de Sitter space/Conformal
Field Theory (AdS/CFT)  conjecture \cite{Maldacena}, such as the strongly
coupling, like type II superconductors \cite{gubser}-\cite{Hart}.
holographic superconductors, Quark-Gluon plasma, and
superconductor/superfluid in condensed matter physics, particularly
using qualitative approaches \cite{HerzogRev}-\cite{Momeni:2014efa}. In  \cite{hee1,hee2}  it was shown that entanglement entropy of any entangled CFT system could be calculated using the area of the minimal surfaces in bulk. This idea called holographic entanglement entropy (HEE) and finds a lot of applications 
 \cite{app1}-\cite{app14}, as
a geometric approach. In order to address this issues, we consider
two possible portions $\tilde{A}$(set A), $B=\tilde{A}'$ (the
complementary set) of a single quantum system. We make a complete Hilbert
space $\mathcal{H}_{\tilde{A}}\times \mathcal{H}_{\tilde{A}'}$. Let us to consider the Von-Neumann entropy
$S_{X}=-Tr_{X}(\rho\log\rho)$ , here $Tr$ is the quantum trace of quantum operator
$\rho$ over quantum basis $X$. If we compute $S_{\tilde{A}}$ and
$S_{\tilde{A}'}$, we obtain
$S_{\tilde{A}}=S_{\tilde{A}'}$. A physical meaning is 
that Von-Neumann entropies are now more likely to identify it with the
surface boundary area $\partial\tilde{A}$ \cite{Srednicki:1993im}. Following to the  proposal \cite{hee1,hee2},
every quantum field theory in $(d)$ dimensional flat boundary  has a gravitational dual embedded
in $AdS_{d+1}$ bulk. The holographic algorithm tells us that the HEE of a region of space $\tilde A$
and its complement from the $AdS_{d+1}$ geometry of bulk:
\begin{eqnarray}
S_{\tilde A}\equiv S_{HEE} = \frac{Area(\gamma_{\tilde A})}{4G_{d+1}},
\end{eqnarray}
We should firstly find the minimal $(d-1)$D mini-super surface $\gamma_{\tilde A}$. It had been
assumed to extend $\gamma_{\tilde A}|_{AdS_{d+1}}$ to bulk, but
with criteria to keep surfaces with same boundary  $\partial
\gamma_{\tilde A}$  and $\partial {\tilde A}$. 
Computation of the HEE have been initiated
\cite{Peng:2014ira}-\cite{Romero-Bermudez:2015bma}.

Another surprising development is that HEE can be used to describe phase transitions in CFT systems. Here we discuss this issue of  observing phase transition in a two dimensional holographic superconductor, and show that in this approach the first order phase transition can be deteced.

\par
%%%%%%%%%%%%%%%%%%%%%%%%%%%%%%%%%%
\section{ Model for 2D HSC} 
A toy model for two dimensional holographic superconductors (2DHSC) using the 
$AdS_3/CFT_2$  proposed by the following action
\cite{Liu:2011fy}-\cite{Nurmagambetov:2011yt}:
\begin{eqnarray}\label{action}
&&S=\int d^3
x\sqrt{-g}\Big[\frac{1}{2\kappa^2}(R+\frac{2}{L^2})\\&&\nonumber-\frac{1}{4}F^{ab}F_{ab}-|\nabla\phi-i
 A\phi |^2-m^2|\phi|^2\Big].
\end{eqnarray}
Here, $\kappa^2=8\pi G_3$ defines the three dimensional gravitational constant
,  the Newton constant $G_3$, $L$ is the AdS radius, $m^2=m_{\phi}^2\in(-1,\infty)$ mass of scalar field, and $g=det(g_{\mu\nu})$.

We adopt the following set of planar coordinates for asymptotically $AdS_3$ spacetime:
\begin{eqnarray}\label{g}
ds^2=-f(r)e^{-\beta(r)}dt^2+\frac{d\gamma^2}{f(r)}+\frac{r^2}{L^2}dx^2~.
\end{eqnarray}
To have thermal behavior we choose the temperature for $CFT_2$ from our $AdS_3$ black hole following Bekenstein-Hawking formula:
\begin{eqnarray}
&&T=\frac{f'(r_{+})e^{-\beta(r_+)/2}}{4\pi}.
\end{eqnarray}
here $r_{+}$ is the horizon the the blackhole , the largest root of the Eq. $f(r_{+})=0$.
To preserve staticity, we select the that the Abelien gauge field $A_{\mu}$ and scalar field $\phi$ are static (time independent) and spherically symmetric :
\begin{eqnarray}
A_t=A(r)dt,\ \ \phi\equiv\phi(r).
\end{eqnarray}
It is useful to write the set of the equations of motion (EOM)s in terms of the new radial coordinate
 $z =
\frac{r_+}{r}$,. Using this coordinate the black hole horizon shifted to the $z=1$ and AdS boundary locates at $z=0$. We need to study solutions of the following set of non linear ordinary differential Eqs. near $z=0$ :
\begin{eqnarray}
&&\phi ''+\frac{\phi '}{z}
\left[1+\frac{zf'}{f}-\frac{z\beta
'}{2}\right]+\frac{r_{+}^2\phi}{z^4} \left[\frac{
A^2 e^{\beta
}}{f^2}-\frac{m^2}{f}\right]=0~,\label{phiz}\\&&
A ''+\frac{A '}{z} \left[1-\frac{z\beta '}{2}\right]-\frac{2 r_{+}^2 A \phi ^2}{z^4f}=0~,\label{Az}\\&&
\beta '-  \frac{4 \kappa ^2r_{+}^2}{z^3} \left[\frac{A
^2 \phi^2 e^{\beta}}{f^2}-
\frac{z^4\phi
'^2}{r_{+}^2}\right]=0\label{betaz},
\\&&
f'-\frac{2 r_{+}^2}{L^2z^3}- \kappa ^2 z
e^{\beta} A'^2-\frac{2 \kappa ^2 m^2 r_{+}^2\phi ^2}{z^3}
\\&&\nonumber-\frac{2 \kappa ^2 r_{+}^2}{z^3} \left[\frac{ A
^2 \phi^2 e^{\beta}}{f}+\frac{f \phi
'^2z^4}{r_{+}^2}\right]=0,\label{fz}
\end{eqnarray}
The numerical and analytical solutions of the above Eqs. with $\phi\neq0, T<T_c$ were presented in Ref. \cite{Liu:2011fy,Momeni:2013waa}. \par
%%%%%%%%%%%%%%%%%%%%%%%%%%%%%%%%%%%
\section{Calculation of holographic entanglement entropy} 
We need to specify the bulk system to find minimal area surfaces. The appropriate 
 systems  parametrization id to represent bulk sector in one
degree of freedom $\tilde A:=\{t=t_0,-\theta_0\leq
\theta\leq\theta_0,r=r(\theta)\}$. Using metric given in Eq. (\ref{g}),
 the total length (angle) $\theta_0$  and HEE are defined by the following integrals:
\begin{eqnarray}
\label{theta0}&&\theta_0=\int_{0}^{\theta_0}{d\theta}=\int_{0}^{\theta_0}{\frac{C dr}{r\sqrt{f(r)(r^2-C^2)}}}\\&&
\label{shee}S_{HEE}\equiv \frac{1}{2G_3}\int_{0}^{\theta_0}{\frac{rdr}{\sqrt{f(r)(r^2-C^2)}}}
\end{eqnarray}
We need to evaluate the sensitivity of
(\ref{theta0},\ref{shee}) in the bulk of acute regimes of
temperature.We apply
 the domain wall
approximation analysis \cite{Albash2012} which is based on the idea is
proposed to investigate some aspects of the HEE along renormalization
group (RG) trajectories.  The $AdS_3$ is relating to RG flows in $(1+1)D$.  \footnote{The RG flow is defined as the $N=1$ SUSY
deformation of $N=4$ SUSY-YM theory.} The geometry (metric) which we
will use is called  here domain wall geometry, in which we suppose that two pats of the AdS are connected by a wall with two AdS radius $L_{IR}>L_{UV}$.
Furthermore we suppose that the length scale of AdS $L$, is defined
as the following:

\begin{equation}
L = \left\{ \begin{array}{lr}
L_{UV} \ , & r > r_{DW} \\
L_{IR}\ , & r < r_{DW}
\end{array} \right. \ .
\end{equation}
%%%%%%%%%%%%%%%%%%%%%%%%%%%%%
We have a sharp phase transition between two patches of the
AdS space time, so we locate the DW at the radius  $r=r_{DW}$ . In this approach we define 
$r_*$ as a possible ``turning'' point of the minimal surface $\gamma_{\tilde A}$. It is defined by $r'(\theta)|_{r=r^{*}}=0$. We will suppose that $r_*<r_{DW}$.. We replaced the integrating out to  $r=+\infty$ by integrating out to large positive radius $r_{UV}$.  Indeed,  we assume that $r_{UV}$
stands out for  UV cutoff
\cite{Romero-Bermudez:2015bma}.
We can rewrite HEE as we like:

\begin{eqnarray}
&&S_{HEE}=\frac{1}{2G_3}\Big[S_{IR}+S_{UV}\Big],\\&&
\label{SIR}S_{IR}=\int_{r_*}^{r_{DW}}{\frac{rdr}{\sqrt{f_{IR}(r^2-L_{IR}^2)}}},\\&&\label{SUV}
S_{UV}=\int_{r_{DW}}^{r_{UV}}{\frac{rdr}{\sqrt{f_{UV}(r^2-L_{UV}^2)}}}.
\end{eqnarray}
In both cases IR,UV, the geometry of AdS has imposed tight constraints on the metric. Furthermore the angle is defined by a two partion integral, 
 $\theta_{\text{IR}},\theta_{\text{UV}}$ :
\begin{eqnarray}
&&\theta_{IR}=i\log\Big[\frac{r_{*}}{r_{DW}}\frac{\sqrt{L_{IR}^2-r_{DW}^2}-L_{IR}}{\sqrt{L_{IR}^2-r_{*}^2}-L_{IR}}\Big],\\&&
\theta_{UV}=\frac{\sqrt{r_{UV}^2-L^2}}{r_{UV}}-\frac{\sqrt{r_{DW}^2-L^2}}{r_{DW}}.
\end{eqnarray}
The entangelement entropy can be computed  as the following form:
\begin{eqnarray}
&&S_{IR}=\sqrt{r_{DW}^2-L_{IR}^2}-\sqrt{r_{*}^2-L_{IR}^2}
\\&&S_{UV}=\frac{iL^2}{L_{UV}}\log\Big[\frac{r_{DW}}{r_{UV}}\frac{\sqrt{L_{UV}^2-r_{UV}^2}-L_{UV}}{\sqrt{L_{UV}^2-r_{DW}^2}-L_{UV}}\Big]
\end{eqnarray}
\par
%%%%%%%%%%%%%%%%%%%%%%%%%%%%%%%%%%%%%%%%%%%%%%%
\section{HEE close to the  $T \lesssim T_c$ }

 The normal  phase of the system can even be achieved when we set the scalar field $\phi=0$. In this case the  EE between $\tilde{A}$ and its complement is given by:
\begin{eqnarray}
&&s_{\tilde{A}}=4 G_3 S_{HEE}=
2r_{*}^{-1}\int_{r_{UV}}^{r_{*}}{\frac{rdr}{\sqrt{f(r)(r^2-r_{*}^2)}}}
\end{eqnarray}
We rewrite  $s_{\tilde{A}}$
in terms of the coordinate $z$:
\begin{eqnarray}
s_{\tilde{A}}=2r_{+}r_{*}\int_{z_{UV}}^{z_{*}}\frac{dz}{z^3\sqrt{f(z)}\sqrt{z^{-2}-z_{*}^{-2}}}.
\end{eqnarray}
Near the   critical point  $T \lesssim T_c$ we can approximate the integral as the following:
\begin{eqnarray}
s_{\tilde{A}}=2r_{+c}r_{*}\int_{z_{UV}}^{z_{*}}\frac{dz}{z^3\sqrt{f_0}\sqrt{z^{-2}-z_{*}^{-2}}}\label{s-Tc}.
\end{eqnarray}
and for the angle,
\begin{eqnarray}
&&\frac{\theta_0}{2}=r_{*}\int_{r_{UV}}^{r_{*}}\frac{dr}{r\sqrt{f(r)(r^2-r_{*}^2)}}\\&&\nonumber=
\frac{r_{*}}{r_{+}}\int_{z_{UV}}^{z_{*}}\frac{dz}{z\sqrt{f(z)(z^{-2}-z_{*}^{-2})}}.
\end{eqnarray}
At criticality when  $T \lesssim T_c$ we obtain:
\begin{eqnarray}
&&\frac{\theta_0}{2}=
\frac{r_{*}}{r_{+c}}\int_{z_{UV}}^{z_{*}}\frac{dz}{z\sqrt{f_0(z^{-2}-z_{*}^{-2})}}\label{theta-Tc}.
\end{eqnarray}
where
\begin{eqnarray}
T_c=\frac{1}{4\pi r_{+c}}\Big(2r_{+c}^2L^{-2}-\kappa^2\mu_c^2\Big)\label{Tc}
\end{eqnarray}
We evaluate the  HEE and angle(length) numerically with $T_c=0.01$.
Numeric analysis showed \cite{Momeni:2015iea}:
\begin{itemize}
\item 
When we increase the temperature and dual chemical potential, HEE is also increasing smoothly. System becomes a mormal conductor at high temperatures and tends to a superconductor phase when temperature decreases. 

\item If we consider HEE in constant temperature (isothermal curves), we showed that there exists 
at least one lower temperature
regime  $T<T_c$ is almost compulsory for superconductivity. Furthermore, we conclude that there is 
no "confinement/deconfinemnet" phase transition point  in the
$CFT_2$.
\item If we fixed the relative chemical potential 
 $\frac{\mu}{\mu_c}$, we observe that by increasing temperature we have much more amount of HEE. The slope 
of the HEE  $\frac{dS}{dT}$
decreases as the relative chemical potential
$\frac{\mu}{\mu_c}\neq1$ decreases. We interpret this phenomena as emergent of more 
degrees of freedom (dof) 
at low temperature regimes. 
\item  Both of the HEE and the characterestic length of the entangled system 
are always increasing with respect to the temperature $T$, and
never decreasing for fixed $T_c$.

\item HEE is linear function of 
 length. One reason is that in small values of belt angle (small sizes) the system emerges
new extra dof. Another reason could be underestood via first law of entanglememt entropy
\cite{Bhattacharya:2012mi},\cite{Momeni:2015vka}.

\item When we add the scalar field to the system, 
$\phi(z)\neq0$,if we adjust data as
$\epsilon_0^2 = 0.05,\kappa \mu_c=0.005$., we found the critical temperature
 $T_c=0.2$. It was demonstrated that the system softly has transition from normal phase
$T>T_c$ to the superconductor phase $T \lesssim T_c$ for $T\approx
0.0179, 0.0173, 0.0165, 0.0152, 0.0132$. 
\item 
We demonstrated that 
 the slope of the HEE with respect to the belt angle
$\frac{ds'}{d\theta'}$ remains constant. It means that there is no critical belt length $\theta_c$ in our $CFT_2$ system. 

\end{itemize}

\section {First order phase transition at critical point }

A numerical study
of HEE and length shows that these quantities were smooth,
and that their behaviors went most smoothly when the phase
transition held the same mechanism as the usual. When the temperature of the system
$T$ rattends to its critical value $T_c$ , we detect a discontinuity  in
the $\frac{d S}{dT}$  when the phase is changed, and a  first order
phase transition may be introduced into the system. By computing the difference
between $\frac{d S}{dT}$ for $T > T_c$  and $T< T_c
$  at $T=T_c $ (scaled the
entropy in $\log$ scale) we observe that this phase transition is  of the  first order.  Indeed, at the
critical point $\lim \frac{d S}{dT}|_{T\to T_c}=\infty$ and we observe the  first order
phase transitions from the behavior of the entanglement entropy
HEE at the critical point $T=T_c$.  These types of first order
phase transitions have been observed recently in literature
\cite{Momeni:2015vka}. We conclude that the HEE  is indeed a good probe
to phase transition in lower dimensional holographic
superconductors. Furthermore, it implies that the HEE can indicate
not only the occurrence of the phase transition, but also we can
learn about  the order of the phase transition from it.

\par
\section{Summary} 

Prior work has documented the effectiveness of holographic entanglement entropy in describing phase transition of quantum systems. Liu et al. \cite{Liu:2011fy} for example, reports that holographic superconductors in two dimensional systems moving towards the phase transition in lower dimensional quantum field theory systems improved across several studies.  However, these studies have either been purely bulk studies or have not focused on thermodynamic whose entanglement entropy was boundary-related. In this study we showed that entanglement entropy predicts a first order phase transition in two dimensional holographic superconductors.


\begin{thebibliography}{10}
\bibitem{Tameem Albash}
We are indebted to Tameem Albash for pointing out this comment.
\end{thebibliography}
\end{document}